\documentclass[twocolumn,flushbottom,showpacs,preprintnumbers,superscriptaddress,amsmath,amssymb]{revtex4}
\usepackage{graphicx}
\usepackage{dcolumn}
\usepackage{bm}
\usepackage{amssymb}
\usepackage{amssymb}
\usepackage{color}
\def\>{\rangle}

\makeatletter
\newcommand{\rmnum}[1]{\romannumeral #1}
\newcommand{\Rmnum}[1]{\expandafter\@slowromancap\romannumeral #1@}
\makeatother

\begin{document}

\title{Noise suppression of on-chip mechanical resonators by chaotic coherent feedback}

\author{Nan Yang}
\affiliation{Department of Automation, Tsinghua University,
Beijing 100084, P. R. China} \affiliation{Center for Quantum
Information Science and Technology, TNList, Beijing 100084, P. R.
China} \affiliation{CEMS, RIKEN,
Saitama, 351-0198, Japan}
\author{Jing Zhang}\email{jing-zhang@mail.tsinghua.edu.cn}
\affiliation{Department of Automation, Tsinghua University,
Beijing 100084, P. R. China} \affiliation{Center for Quantum
Information Science and Technology, TNList, Beijing 100084, P. R.
China} \affiliation{CEMS, RIKEN,
Saitama, 351-0198, Japan} \affiliation{State Key Laboratory of
Robotics, Shenyang Institute of Automation Chinese Academy of
Sciences, Shenyang 110016, China}
\author{Hui Wang}
\affiliation{Institute of Microelectronics, Tsinghua University,
Beijing 100084, P. R. China} \affiliation{Center for Quantum
Information Science and Technology, TNList, Beijing 100084, P. R.
China}
\author{Yu-xi Liu}
\affiliation{Institute of Microelectronics, Tsinghua University,
Beijing 100084, P. R. China} \affiliation{Center for Quantum
Information Science and Technology, TNList, Beijing 100084, P. R.
China}
\author{Re-Bing Wu}
\affiliation{Department of Automation, Tsinghua University,
Beijing 100084, P. R. China} \affiliation{Center for Quantum
Information Science and Technology, TNList, Beijing 100084, P. R.
China}
\author{Lian-qing Liu}
\affiliation{State Key Laboratory of Robotics, Shenyang Institute
of Automation Chinese Academy of Sciences, Shenyang 110016, China}
\author{Chun-Wen Li}
\affiliation{Department of Automation, Tsinghua University,
Beijing 100084, P. R. China} \affiliation{Center for Quantum
Information Science and Technology, TNList, Beijing 100084, P. R.
China}
\author{Franco Nori}
\affiliation{CEMS, RIKEN,
Saitama, 351-0198, Japan} \affiliation{Physics Department, The University
of Michigan, Ann Arbor, Michigan 48109-1040, USA}

\date{\today}

\begin{abstract}
We propose a method to decouple the nanomechanical resonator in
optomechanical systems from the environmental noise by
introducing a chaotic coherent feedback loop. We find that the
chaotic controller in the feedback loop can modulate the dynamics
of the controlled optomechanical system and induce a
{broadband} response of the mechanical mode. This
{broadband} response of the mechanical mode will
cut off the coupling between the mechanical mode and the
environment and thus suppress the environmental noise of the
mechanical modes. As an application, we use the protected
optomechanical system to act as a quantum memory. {It's shown} that the {noise-decoupled} optomechanical
quantum memory is efficient for storing information transferred
from coherent or squeezed light.
\end{abstract}

\pacs{03.67.Pp, 02.30.Yy}
\maketitle

\section{\label{sec:level1}Introduction\protect\\ }

Optomechanical systems have attracted intense {attention} in recent
years due to its extensive applications~\cite{OM1,OM2,OM3,OM4},
and rapid {progress has} been made both theoretically and
experimentally in related
fields{~\cite{OM5,OM6,OM7,OM8,OM9,OM10,OM11,QE1,QE2,QE3,QE4,GD1,GD2,QM1,QM2}}.
One of the most interesting {problems} for
optomechanical systems is to explore the quantum
{aspects} of mechanical
motion~\cite{QE1,QE2,QE3,QE4}, which is important not only for fundamental
studies of quantum mechanics, but also for further applications,
such as the detection of gravitational {waves}~\cite{GD1,GD2},
{and} quantum memorise~\cite{QM1,QM2}. However, these quantum effects will
be damaged by environmental noises. {Although}
the recent development of experimental techniques
{have} made it possible to cool mechanical modes
to the ground state~\cite{QE1,CE1,CE2,CE3,CE4}, the mechanical
quantum superposition state is still too fragile under
environmental noises, and thermal noise will be dominant if
the mechanical mode is far from the ground state.

Due the problems mentioned above, how to suppress the
environmental {noises} more efficiently is crucial in exploring the
quantum-classical boundary of {nanomechanical resonators}. One possible way to solve this problem is to
introduce {either} active or passive feedback to compensate the noise
effects{~\cite{CE1,CE2,CE3,CE4,CS1,CW1,FC1,FC2,FC3,CC1}}.
{Side band cooling{~\cite{CE1,CE2,CE3,CE4,CS1,CW1}} is the most
widely-used passive compensation method, and
experiments~\cite{CE1,CE2,CE3,CE4} in both {the strong}
 and {the weak optomechanical coupling}
regimes have been reported to reach} the quantum-mechanical ground state~\cite{CE1,CE2,CE3,CE4}.
Approaches based on active feedback
compensation{~\cite{FC1,FC2,FC3,CC1}}{,} are also effective in suppressing{
 environmental noise}. The essence of these methods is to steer the system to the
desired state {by} using the measurement output from a particular quantum
nondemolition measurement. Another possible way to solve this
problem is to decouple the mechanical resonator from the heat bath
by introducing a carefully-designed open-loop control. Dynamic
decoupling control (DDC)~\cite{DD1} and its optimized
versions~\cite{DDO1,DDO2,DDO3,DDO4} are possible ways to achieve
this, which introduce high frequency control {pulses} to average out
the low frequency noises. However, it is not easy to generate the
required high-frequency or optimized pulse in optomechanical
systems and{,} thus{,} to our knowledge{,} DDC has never {been} used to protect
the mechanical states in such systems.

Motivated by the DDC-type control and especially our recent
work~\cite{DC1} (introducing a broadband chaotic control to
suppress decoherence of a superconducting qubit~\cite{DC1}), in
this paper, we propose a method to decouple the {nanomechanical}
resonator from {its} environmental {noises} by introducing a chaotic
coherent feedback loop. Based on the theory of coherent
feedback~\cite{CF1,CF2,CF3,CF4,CF5,CF6,CF7,CF8,CF9,CF10,Jacobs},
which is one of the major quantum feedback
approaches~\cite{QFC1,QFC2,QFC3,QFC4}, the basic idea of our
method is to transfer {a} broadband chaotic control signal from
the controller to the controlled optomechanical systems by
feedback {connections}. This broadband control {induces} an effective
broadband frequency shift of the mechanical resonator and then {decouples} the mechanical mode from the environmental noises.
{Afterwards}, we use the protected mechanical mode as a quantum
memory to store
continuous-variable quantum signals, such as coherent states and
squeezed states, which may have potential applications.

This paper is organized as follows. In {Sec.~\Rmnum{2}}, we {will provide}
general discussions to show the noise-decoupling mechanism for our
chaotic feedback strategy. The possible physical implementations
for our {noise-decoupling} strategy in on-chip optomechanical systems {are}
discussed in {Sec.~\Rmnum{3}}. As an application, in {Sec.~\Rmnum{4}}, we show
how to use a optomechanical system, protected by the designed
chaotic feedback control, to act as a quantum memory. In {Sec.~\Rmnum{5}},
we {summarize} the conclusions and {provide} a few forecasts of future work.

\section{\label{sec:level1}noise decoupling by chaotic feedback\protect\\}

In this section, we show the mechanism of our
chaotic-feedback-induced noise decoupling strategy, in particular for
quadratically-coupled optomechanical
systems~\cite{QCS1,QCS2,QCS3,QCS4,QCS5}. This is motivated by our
previous work~\cite{DC1} which shows that decoherence in
supercoducting circuits can be greatly suppressed by chaos which is typically believed to be a source of decoherence. The main idea of the
chaos-induced decoherence suppression approach is to introduce a
broadband chaotic signal to "randomly" kick the system and
compensate the effects of noise. This is somewhat similar to the
noise suppression approaches by the quantum Zeno effect in which
random signals are introduced to kick the system to compensate the
noise effect. However, chaotic signals are deterministic
signals and thus will not introduce additional decoherence.

Note that, there are some difficulties in introducing such kind of
chaotic control to suppress the noises of the quantum-mechanical mode in
optomechanical systems: {(\rmnum{1})} it is quite hard to drive the
mechanical mode of an optomechanical system directly by a chaotic
acoustic field; and {(\rmnum{2})} the optical cavity in the optomechanical
system will work as a low-pass filter to squeeze the broadband
chaotic signal if we drive the system directly by an open-loop
chaotic optical signal and thus make the control signal not so
"random", which would lead to a failure of our
decoherence-suppression approach. To solve these problems, we
introduce a particular coherent feedback
loop to break the symmetry of the
optomechanical system. Thus, the chaotic controller in the
feedback loop can broaden the bandwidth and preserve the
high-frequency components of the mechanical mode, and protect
it from the environmental noises.



\begin{center}
\begin{figure}[ht]
\centering
\includegraphics[width=\textwidth,bb=0 0 750 260]{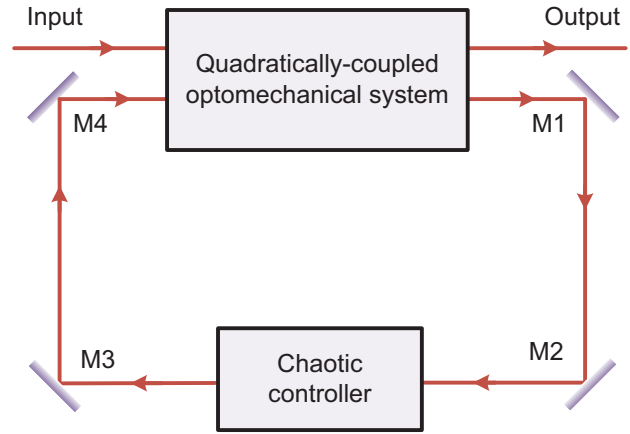}
\caption{ (Color online) Schematic diagram of the noise-decoupling
system by coherent feedback modulation. Two quantum components,
i.e., a quadratically-coupled optomechanical system and a chaotic
controller, are connected by {the} mediated optical
{fields}. The output of the optomechanical system
is taken as the input fed into the chaotic controller.
{Also,} the chaotic signal generated by the chaotic
controller is then fed back to control the dynamics of the
quadratically-coupled optomechanical system. The M1, M2, M3, M4
represent total-reflection mirrors that are introduced to change
the light path. In this model, the radiation pressure can directly
change the frequency of the mechanical resonator because the
coupling between the mechanical mode and the optical mode is
quadratic.} \label{decouple}
\end{figure}
\end{center}

As illustrated in Fig.~\ref{decouple}, our
feedback control system consists of two components, i.e., a
quadratically-coupled optomechanical system~(the controlled
system) and a chaotic controller. These two components are
connected by a mediating optical field, from which we can construct a
field-mediated coherent feedback system~\cite{CF1,CF2,CF3,CF4,Jacobs}. In
the interaction picture of the noise frequency $\omega$, the Hamiltonian of the controlled device,
i.e., the quadratically-coupled optomechanical device, can be
written as~\cite{QCS1,QCS2,QCS3,QCS4,QCS5}
\begin{eqnarray}\label{H_1}
H_1&=&\omega_{a_1} a_1^\dag a_1 + G_1 a_1^\dag a_1 b_1^\dag
b_1 + \Omega_{1} b_1^\dag b_1\nonumber\\
&&+i\varepsilon_1\left[a_1^\dag \exp({-i\omega_{d_1}t})-a_1
\exp({i\omega_{d_1}t})\right]\nonumber\\
&&+\sum_{\omega}g(\omega)\left[b^\dag(\omega) b_1 e^{-i\omega
t}+b(\omega) b_1^{\dag} e^{i\omega t}\right],
\end{eqnarray}
where $a_1$ and $b_1$ denote the annihilation operators of the
cavity {mode} and the mechanical mode in the quadratically-coupled
optomechanical system, and $\omega_{a_1}$, $\Omega_{1}$ are the {natural}
frequencies of these two modes. Here, we assume that $\hbar=1$.
The optomechanical coupling we consider here is a kind of
quadratic optomechanical interaction with strength
$G_1$~\cite{QCS1,QCS2,QCS3,QCS4,QCS5}. The optical mode $a_1$ is driven
by an external driving field with strength $\varepsilon_1$ and
 frequency $\omega_{d_1}$. Here $b(\omega)$ represents the noise
mode with frequency $\omega$ acting on the mechanical mode and
$g({\omega})$ is the coupling strength between the mechanical mode
and the noise mode.

Here we use $H_c$ to denote the Hamiltonian of the chaotic
controller, and $a_2$ denotes the annihilation operator of the
chaotic cavity field in the controller. Then the interaction
Hamiltonian of the quadratically-coupled system and the controller
$H_{\rm{int}}$ takes the form (see Appendix A)
\begin{equation}\label{eq.2}
H_{\rm{int}}=\frac{1}{2i}(\sqrt{\gamma_1 \gamma_2}-\sqrt{\gamma_2 \gamma_f})(a_2^\dag a_1-a_1^{\dag} a_2),
\end{equation}
where $\gamma_1$ and $\gamma_2$ represent the damping rates of the
optical cavities in the controlled system "1" and the chaotic
controller "2", and $\gamma_f$ denotes the damping rate of the controlled
cavity induced by the feedback field. {The} total
Hamiltonian of the coherent feedback loop is provided by
\begin{equation}\label{H_tot}
H_{\rm{tot}}=H_1 + H_c +H_{\rm int}.
\end{equation}

In the strong-driving regime, the optical fields in the
quadratically-coupled optomechanical system and the chaotic
controller can be treated classically. Here we replace the
operator $a_1$ by $\alpha_1(t)$, which represents the
{classical part} of the optical field $a_1$, and {then} eliminate the classical
parts including $H_c$ and $H_{\rm int}$ in the total Hamiltonian.
Thus the Hamiltonian of the feedback control system given in
Eq.~(\ref{H_tot}) can be simplified as
\begin{equation}
\begin{split}\label{H_effective}
H_{\rm eff}=&\Omega_{1}{b_1} b_1^\dag + f(t)\;b_1^\dag b_1
\\
&+ \sum_{\omega}g({\omega})\left[{b({\omega}) b_1^\dag \exp({i\omega t})+ \rm {h.c.}} \right],
\end{split}
\end{equation}
where $f(t)=G_1 {|\alpha_1(t)|}^2$, and the amplitude of the cavity
field {$|\alpha_1(t)|$} is modulated by the chaotic
controller and thus it is a broad-band signal. The effective Hamiltonian
in Eq.~(\ref{H_effective}) includes three parts:
{(\rmnum{1})}~the free Hamiltonian of the
mechanical mode with {natural} frequency $\Omega_1$;
{(\rmnum{2})}~a correction term with the mechanical
frequency shift $f(t)$ induced by the chaotic controller {$H_c$};
{(\rmnum{3})}~the interaction Hamiltonian {$H_{\rm int}$} between
the mechanical mode {$b_{1}$} and {its} environmental noises {$b(\omega)$}.
In the rotating reference frame with {the} unitary operator
\begin{equation}\label{eq.11}
U=\exp \left[-{i}\int^{t}_{0}{\left(f(\tau)+\Omega_1\right)b_1^\dag b_1}d\tau \right],
\end{equation}
the effective Hamiltonian {is given by}
\begin{eqnarray}\label{H_effective1}
\tilde{H}_{\rm eff}&=&U^\dag H_{\rm eff}
U-iU^{\dag}\partial{U}/\partial{t}\nonumber\\
&=&\sum_\omega {g}(\omega) \left[ b(\omega) b_1^\dag
e^{{-i(\Omega_1-\omega)t}-i\int_{0}^{t}{f(\tau)}d\tau} + {\rm
h.c.}\right].\nonumber\\
\end{eqnarray}
By averaging over the {broadband} signal $f(t)$~\cite{APS1}, we
have (see Appendix B)
\begin{equation}\label{tg}
\overline{\exp\left[-i\int_{0}^{t}{f(\tau)}d\tau\right]}=\sqrt{M},
 \end{equation}
where $M$ is a correction factor. Thus, {the effective Hamiltonian shown in }Eq.~(\ref{H_effective1})
can be simplified as
\begin{equation}\label{eq.11}
\tilde{\tilde{H}}_{\rm eff}= \sum_{\omega} {\tilde{g}}({\omega})\left\{b({\omega}) b_1^\dag \exp[{-i(\Omega_1-\omega)t}]+ {\rm h.c.} \right\},
\end{equation}
where $\tilde{g}({\omega})=\sqrt{M}g({\omega})$ is the modified
coupling strength between the mechanical mode and the heat bath
after introducing the chaotic signal $f(t)$. It can be seen that
the modified coupling strength $\tilde{g}(\omega)$ can be greatly
decreased if the correction factor $M$ is small enough, under
which the mechanical mode is efficiently decoupled from the
environmental noises.

As shown in Appendix B, the correction factor $M$ is determined by
the power spectrum $S_{f}(\omega)$ of the chaotic signal $f(t)$
\begin{equation}\label{M}
M=\exp\left[-\pi\int^{\omega_u}_{\omega_l}\frac{S_{f}(\omega)}{\omega^2}d\omega \right],
\end{equation}
where $\omega_u$ and $\omega_l$ are the upper bound and lower bound of the frequency band of the chaotic signal $f\left(t\right)$. Note that, $M$ varies from $0$ to $1$. Specially, $M=0$ corresponds to the full-decoupling case, and $M=1$ corresponds to the case without decoupling. {Since} the power spectrum $S_{f}(\omega)$ is broadened by the chaotic modulation, the value of $M$ is thus very small and the mechanical mode is decoupled from the environmental noises.


\section{Physical implementation in on-chip optomechanical systems\protect\\}

In this section, we discuss how to physically implement our
{chaotic-feedback-based noise decoupling} strategy in on-chip optomechanical systems.

\subsection{\label{sec:level1}Implementation of the quadratically-coupled optomechanical system\protect\\ }

\begin{center}
\begin{figure}[ht]
\centering
\includegraphics[width=\textwidth,bb=0 0 600 450]{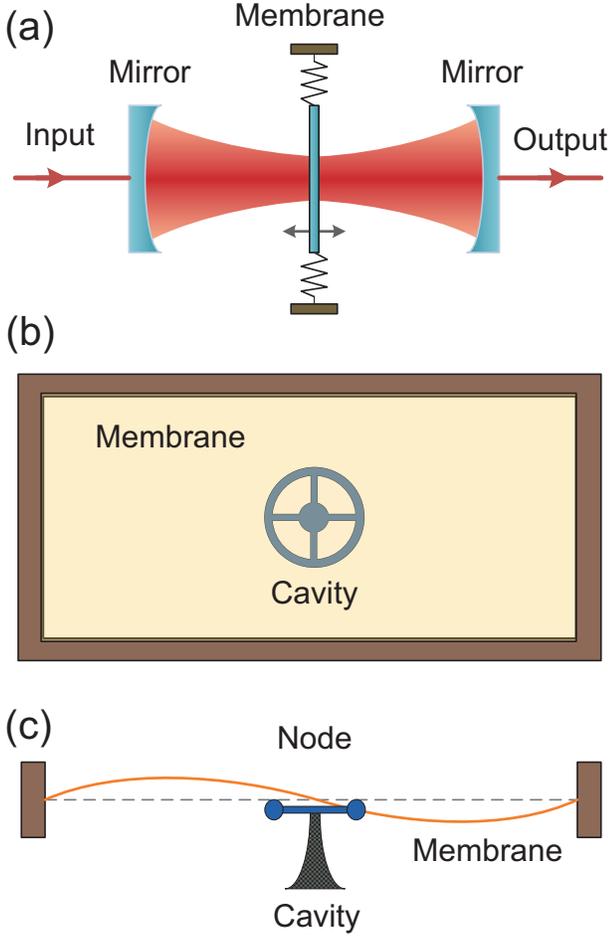}
\caption{(Color online) Schematic diagram of the
quadratic-coupling optomechanical systems with a Fabry-Perot
cavity and the rectangular membrane. (a) Quadratic optomechanical
system with a Fabry-Perot cavity: the quadratic-coupling is realized
by putting a membrane in the middle of the Fabry-Perot cavity. (b)
Top view and (c) cross-sectional view of a rectangular membrane
optomechanical system, where its node {coincides}
with the central point of the cavity. The rectangular membrane
supports various vibrational modes $u=(j,k)$, where $j,k=1,2...$
are the mode indexes. Here the rectangular membrane is driven to
the (1,2) mode, which has two anti-nodes and one node.}
\label{quadratic_coupling}
\end{figure}
\end{center}
Here we list two possible examples of the quadratic-coupling
optomechanical system~\cite{QCS1,QCS2,QCS3,QCS4,QCS5}. The first
example is shown in Fig.~\ref{quadratic_coupling}(a), in which a
membrane is placed in the middle of a cavity and can move freely
under the laser-induced pressure~\cite{QCS1,QCS2,QCS3,QCS4}. Such
kind of structure leads to a quadratic coupling term between the
mechanical mode and the cavity mode. Another example for the
quadratic-coupling is the rectangular membrane optomechanical
system~\cite{QCS5}. As seen in Figs.~\ref{quadratic_coupling}(b) and (c),
the rectangular membrane placed above a toroidal cavity is driven by
the optical field inside the toroidal cavity, which may generate
both linear coupling and quadratic coupling modes between the
cavity field and the membrane. The coupling strengthes of these
two coupling modes are determined by three factors:
{(\rmnum{1})} the vibrational mode of the
rectangular membrane; {(\rmnum{2})} the distance
between the membrane and the upper surface of the toroidal cavity;
and {(\rmnum{3})} the relative position of the
toroidal cavity. Moreover, the coupling modes displayed in the
rectangular membrane optomechanical system can be controlled by
modulating the above factors. The purely quadratic-coupling mode
can be realized when~\cite{QCS5}: {(\rmnum{1})} the
rectangular membrane is excited in a vibrational mode that
contains at least one node; {(\rmnum{2})} the
rectangular membrane is placed right above {the}
toroidal cavity; and {(\rmnum{3})} the node of the
membrane is located at the central point of the cavity.
Under these conditions, the linear coupling term between the
membrane and the cavity field can be completely removed.

The mechanism of the rectangular membrane optomechanical system is
similar to the Fabry-Perot-type quadratic-coupling system, and
they share the same Hamiltonian, which is shown in
Eq.~(\ref{H_1}). Hereafter, we apply our
noise-decoupling method to the rectangular membrane optomechanical
system presented above.

\begin{center}
\begin{figure}[ht]
\centering
\includegraphics[width=\textwidth,bb=0 0 900 350]{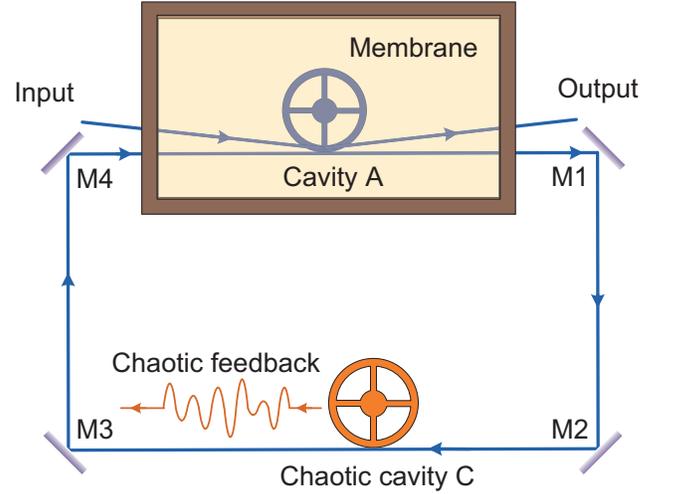}
\caption{(Color online) The noise-decoupling model with the control of a toroidal cavity. Here the toroidal cavity is a chaotic controller, {which} shifts the cavity field of {the} quadratically-coupled optomechanical system to chaos.}
\label{decouple_toroid}
\end{figure}
\end{center}
\subsection{\label{sec:level1}Implementation of the chaotic controller\protect\\ }

In this section, we consider an optomechanical system [see
Fig.~\ref{decouple_toroid}] with chaotic dynamics~\cite{COM1} as
the chaotic controller {in the feedback control
loop}. {For simplicity} we denote the controlled
quadratically-coupled optomechanical device as system $1$, and the
chaotic controller as system $2$. The Hamiltonian of system $1$ is
displayed in Eq.~(\ref{H_1}); and the Hamiltonian of system $2$ is
taken as
\begin{equation}\label{eq.2}
\begin{split}
H_2 =& \omega_{a_2} a_2^\dag a_2 + G_2\,a_2^\dag a_2 (b_2^\dag + b_2) + \Omega_{2}\,b_2^\dag b_2
\\
&+i\varepsilon_2 [a_2^\dag \exp({-i\omega_{d_2}t})-{a_2}\exp({i\omega_{d_2}t})],
\end{split}
\end{equation}
where $a_2$ and $b_2$ denote the annihilation operators of the
cavity mode and the mechanical mode in system $2$; and
$\omega_{a_2}$, $\Omega_{2}$ correspond to their inherent
frequencies. Here $G_2$ denotes the optomechanical coupling
strength in system $2$. {The cavity mode in system 2} is driven by an input laser
field with driving strength $\varepsilon_2$ and corresponding
driving frequency $\omega_{d_2}$. {Here}, the driving
frequencies of the cavity modes in the two systems are chosen to
be: $\omega_{d_1}=\omega_{d_2}=\omega_d$. In the rotating
reference frame with {the} unitary operator
$U=\exp[-i\omega_d(a_1^{\dag}a_1+a_2^{\dag}a_2)t]$, the total
Hamiltonian of the quantum feedback loop can be transformed to the
form
\begin{equation}\label{H_feedback}
\begin{split}
H_{\rm{tot}}=&\Delta_{1} a_1^\dag a_1 + G_1 a_1^\dag a_1 b_1^\dag b_1 + \Omega_{1} b_1^\dag b_1
\\
&+ \Delta_{2} a_2^\dag a_2 + G_2 a_2^\dag a_2 (b_2^\dag + b_2) + \Omega_{2} b_2^\dag b_2
\\
&+i\varepsilon_1(a_1^\dag - {a_1})+i\varepsilon_2 (a_2^\dag-{a_2})
\\
&+\frac{1}{2i}(\sqrt{\gamma_1 \gamma_2}-\sqrt{\gamma_2 \gamma_f})(a_2^\dag a_1-a_1^\dag a_2)
\\
&+ \sum_{\omega}g(\omega)[b^\dag(\omega) b_1 \exp({-i\omega t})+ \rm h.c.],
\end{split}
\end{equation}
where $\Delta_{1}=\omega_{a_1}-\omega_{d}$, and
$\Delta_{2}=\omega_{a_2}-\omega_{d}$, denote the detuning
frequencies of cavities $1$ and $2$. Here $\gamma_1$ and
$\gamma_2$ represent the damping rates of the optical cavities $1$
and $2$, $\gamma_f$ denotes the damping rate induced by the feedback field of the controlled cavity. We use the quantum Langevin
equations to describe the dynamics of the {chaotic} feedback system
\begin{subequations}
\label{Langevin1}
\begin{equation}
\begin{split}
\dot{a}_1=& -i\Delta_1 a_1 - \frac{1}{2}(\sqrt{\gamma_1}+\sqrt{\gamma_f})^{2}a_1-iG_1 a_1 b_1^\dag b_1\\
& -\sqrt{\gamma_2\gamma_f}\,a_2+\varepsilon_1-(\sqrt{\gamma_1}+\sqrt{\gamma_f})a_{1,\rm in},\label{subeq:2}
\end{split}
\end{equation}
\begin{equation}
\begin{split}
\dot{a}_2=&-i\Delta_2 a_2 - \frac{\gamma_2}{2} a_2-iG_2 a_2 (b_1^\dag + b_1)+\varepsilon_2
\\
&-\sqrt{\gamma_1 \gamma_2}\,a_1-\sqrt{\gamma_2}\,a_{2,\rm in},\label{subeq:2}
\end{split}
\end{equation}
\begin{equation}
\dot{b}_1=-i \Omega_{1}b_1-iG_1a_1^\dag a_1 {b_1} - \frac{\Gamma_{\!1}}{2} b_1 - \sqrt{\Gamma_{\!1}}\, b_{1,\rm in},\label{subeq:2}
\end{equation}
\begin{equation}
\dot{b}_2=-i \Omega_{2}b_2-iG_2 a_2^\dag a_2 -
\frac{\Gamma_{\!2}}{2} b_2 - \sqrt{\Gamma_{\!2}}\, b_{2,\rm
in},\label{subeq:3}
\end{equation}
\end{subequations}
where $a_{1,\rm in}$ ($a_{2,\rm in}$) is the input of the optical
cavity in system $1$ ($2$); {$b_{1,\rm in}$ ($b_{2,\rm in}$) and $\Gamma_{\!1}$ ($\Gamma_{\!2}$) are the input and the
damping rate of the mechanical mode in system $1$ ($2$).} We assume that the backaction
of the mechanical mode acting on the optical mode in system $1$ is
very weak, then the evolution of the cavity mode $1$ mainly
depends on Eqs.~(\ref{Langevin1}a), (\ref{Langevin1}b), and
(\ref{Langevin1}d). {In the
strong-driving regime, the semiclassical approximation can be
applied: $a_1=\alpha_1 + \tilde{a}_1$, $a_2=\alpha_2 +
\tilde{a}_2$, and $b_2=\beta_2 + \tilde{b}_2$, where $\alpha_1$,
$\alpha_2$, and $\beta_2$ represent the classical parts and
$\tilde{a}_1$, $\tilde{a}_2$ and $\tilde{b}_2$ denote the
operators for the quantum fluctuations.} {Then} we neglect the
quantum fluctuation terms in Eqs.~(\ref{Langevin1}a),
(\ref{Langevin1}b), and (\ref{Langevin1}d). Thus the evolution of
the classical parts in the total system can be described by
\begin{subequations}
\label{Langevin2}
\begin{equation}
\begin{split}
\dot{\alpha}_1=&-i\Delta_1 \alpha_1 - \frac{1}{2}(\sqrt{\gamma_1}+\sqrt{\gamma_f})^{2} \alpha_1
\\
&\varepsilon_1-\sqrt{\gamma_2 \gamma_f}\,\alpha_2,\label{subeq:2}
\end{split}
\end{equation}

\begin{equation}
\begin{split}
\dot{\alpha}_2=&-i\Delta_2 \alpha_2 - \frac{\gamma_2}{2} \alpha_2-iG_2 \alpha_2 (\beta_2^{\ast} + \beta_2)
\\
&+\varepsilon_2-\sqrt{\gamma_1 \gamma_2}\,\alpha_1,\label{subeq:2}
\end{split}
\end{equation}
\begin{equation}
\dot{\beta}_2=-i \Omega_{1}\beta_2-iG_2\alpha_2^{\ast} \alpha_2 - \frac{\Gamma_{\!2}}{2} \beta_2.\label{subeq:2}
\end{equation}
\end{subequations}

\begin{center}
\begin{figure}[ht]
\centering
\includegraphics[width=3.5in]{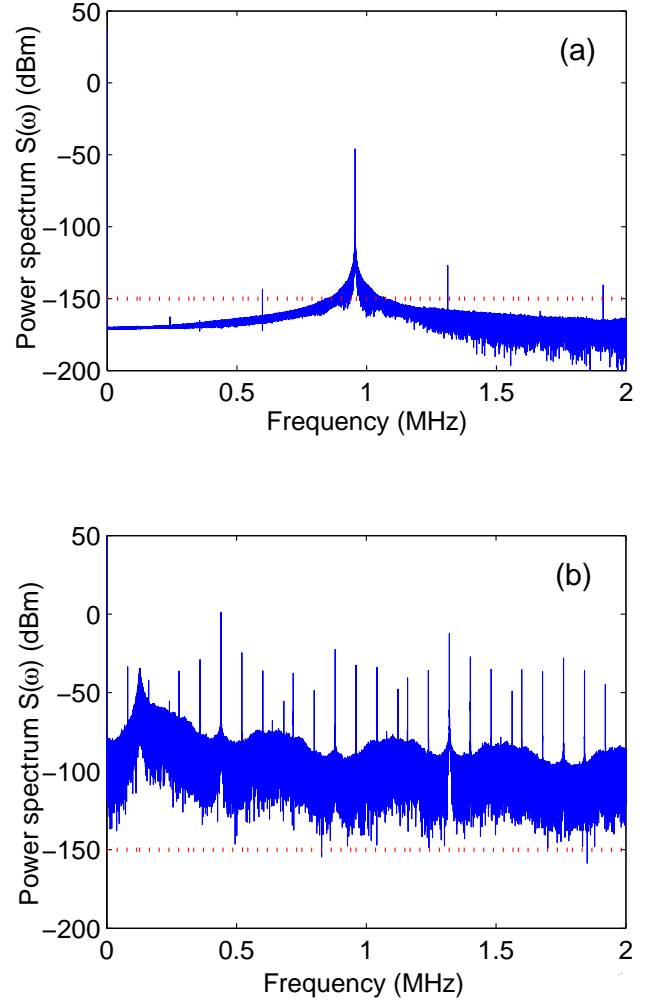}
\caption{(Color online) Power spectra of the mechanical mode in
system $1$: (a) {without feedback} and {(b) with the chaotic feedback is applied}.
(a)~the mechanical membrane is driven by {a} period
signal; (b)~we use a chaotic controller (optomechanical system in
this case) to modulate the power of the mechanical membrane to a wide
region. The parameters are set as follows:
$\Delta_1/{2\pi}=0.75~\rm GHz$,
$\Delta_2/2\pi=0.12~\rm GHz$, $\gamma_1 /2\pi=1~\rm MHz$,
$\gamma_2/2\pi=0.24~\rm GHz$, $\gamma_f/2\pi=0.05~\rm
MHz$, $\Gamma_{\!1}/2\pi=0.01~\rm MHz$, $\Gamma_{\!2}=/2\pi=1.4~\rm MHz$,
$\Omega_1/2\pi=1~\rm MHz$, $\Omega_2/2\pi=0.345~\rm GHz$,
$G_1/2\pi=0.1~\rm MHz$, $G_2/2\pi=0.1~\rm MHz$,
$\varepsilon_1/2\pi=6.6~\rm GHz$, and $\varepsilon_2/2\pi=13.2~\rm
GHz$.} \label{spectra_com}
\end{figure}
\end{center}

When the strength of the driving field $\varepsilon_2$ is strong
enough, the optomechanical system $2$ enters the chaotic regime
and will have a {broadband cavity spectrum}. As the chaotic controller, system $2$ can
spread the spectrum of system $1$ {both in the cavity mode and in the mechanical mode}. Figure.~\ref{spectra_com} shows
the spectrum of the mechanical mode in system $1$ without
[Fig.~\ref{spectra_com}(a)] and with
[Fig.~\ref{spectra_com}(b)] the feedback modulation. As shown in
Fig.~\ref{spectra_com}(a), only a single peak with very small
sidebands is displayed in the spectrum of the mechanical mode if
we do not introduce any feedback modulation. The power of the
background frequency components is very small (less than
-150 dBm). This corresponds to the periodic case. After we
introduce the chaotic feedback [see Fig.~\ref{spectra_com}(b)],
the spectrum of the controlled mechanical mode is greatly
broadened and the whole baseline of the spectrum is increased {to above $150$ dBm}.
This corresponds to the chaotic case, and the broadband response
of the mechanical mode will decouple the mechanical mode from the
environmental noises.

As discussed in {Sec. \Rmnum{2}}, we use the factor $M$ to evaluate the
efficiency of our noise decoupling strategy [see Eq.~(\ref{M})].
The value of $M$ is determined by the spectrum $S_{f}{(\omega)}$
of the signal $f(t)$ (recall that $f(t)=G_1 {|\alpha_1(t)|}^2$), which can be
obtained by numerically solving Eq.~(\ref{Langevin2}). Note that $M\sim1$
when the spectrum $S_{f}{(\omega)}$ is concentrated in a narrow
region, and $M$ will be close to zero if the spectrum
$S_{f}\left(\omega\right)$ is {broadened} by the chaotic modulation. In
our numerical simulations, we find that $M\approx1$ if we do not
introduce feedback [Fig.~\ref{spectra_com}(a)] and $M=0.0074$ if
we introduce the chaotic feedback [Fig.~\ref{spectra_com}(b)],
which coincides with what we expect.

\section{Storage of continuous-variable quantum information}

\begin{center}
\begin{figure}[ht]
\centering
\includegraphics[width=\textwidth,bb=0 0 700 200]{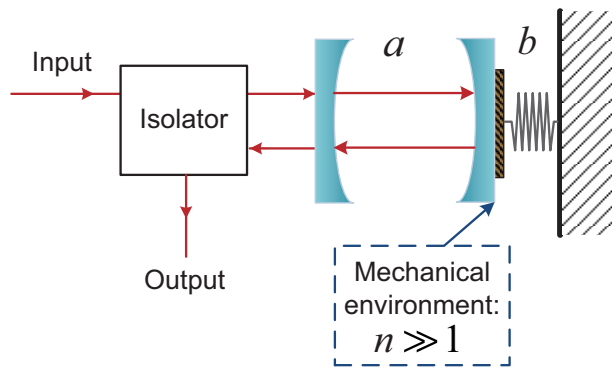}
\caption{(Color online) Schematic diagram of an optmechanical
system for quantum information transfer and storage. A beam of light
with {a} desirable quantum state is fed into a cavity, and then
transferred to the mechanical resonator. {Here $a$ is the cavity mode and $b$ denotes the mechanical mode, $n$ represents the mean thermal excitation phonon
number which follows the Boltzmann distribution.}}
\label{Visio_quantum_transfer}
\end{figure}
\end{center}

The storage of continuous-variable quantum information, i.e., to
realize continuous-variable quantum
memory~\cite{QM1,QM2,QM3,QM5,QM7,QM8}, is important for
quantum communications and quantum computation. One possible way to
solve this problem is to transfer the continuous-variable
information in the optical signal to an on-chip mechanical
resonator which has a lower damping rate. The continuous-variable
optomechanical quantum memory system we consider here is presented
in Fig.~\ref{Visio_quantum_transfer}, which {includes} the input
(output) fields, an optical cavity, and a mechanical
resonator~\cite{QM2}. {By exchanging states, between the cavity mode and the mechanical mode, a quantum state carried by the input field can be written into and stored in the nanomechanical resonator.}

However, the quantum information stored in the mechanical
resonator will unavoidably be destroyed due to the coupling
between the mechanical resonator and the environmental {noise.} Thus, to realize such
kind of continuous-variable quantum memory, we have to suppress
the decoherence effects of the mechanical mode induced by the
environmental noise. As we have discussed in the previous
sections, introducing a chaotic coherent feedback loop to drive
the mechanical mode into the broad-band regime is an efficient way to
decouple the mechanical mode from the environmental noise. In
this section, we will show how to use this noise-decoupled
nano-mechanical resonator as a quantum memory.

Our purpose here is to use a noise-decoupled mechanical resonator
to store continuous-variable information. The key point is how to
transfer a quantum state to a mechanical mode and decouple {this}
mechanical mode simultaneously. Here we propose a strategy with
two optical cavities sharing the same mechanical resonator but
with different optomechanical coupling: one is with linear
optomechanical coupling used for quantum memory; and the other is
with a quadratic optomechanical coupling, used for noise decoupling.
\begin{center}
\begin{figure}[h]
\centering
\includegraphics[width=\textwidth,bb=-10 -10 860 650]{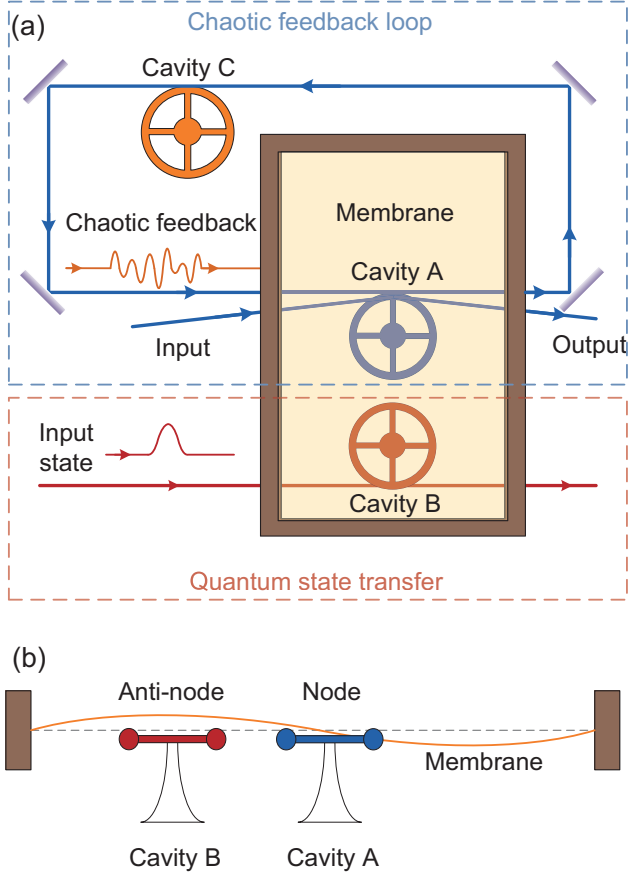}
\caption{ (Color online) (a)~Top view of the quantum memory
system. The noise-decoupled quantum memory system can be divided
into two parts shown by the dashed rectangular grid frames: the chaotic feedback loop
(inside the blue frame) for the noise decoupling of the rectangular
membrane; the setup used for transferring the quantum state
(the red frame) from the input light to the noised-decoupled
rectangular membrane. (b)~Cross-sectional side view of the rectangular membrane
optomechanical system. Cavity A is placed {at} the node of the
rectangular membrane, and cavity B is placed {at} the anti-node.}
\label{quantum_memory}
\end{figure}
\end{center}

Let us now consider how to apply this quantum memory model in {the}
rectangular membrane optomechanical system proposed in
Ref.~\cite{QCS5}. Figure \ref{quantum_memory}(a) shows two toroidal
cavities (A and B) connected to a rectangular membrane. The types
of coupling between the cavity mode and the mechanical mode are
determined by the positions they are placed: the node of the
membrane corresponds to a linear coupling and the anti-node
corresponds to a quadratic coupling
[Fig.~\ref{quantum_memory}(a)]. Thus, we place the toroidal cavity
(cavity A) used for noise decoupling at the node of the membrane;
and the other toroidal cavity (cavity B), used for quantum memory,
{at} the anti-node. Toroidal cavity A is modulated by the chaotic
controller (toroidal cavity C), which leads to the decoupling
between the membrane and its environmental noises. The cavity B is
used for storing the quantum state in the membrane. {The coupling between the
cavity mode and the mechanical mode is assumed to be linear under
the strong-driving regime~\cite{QM1,QM2}.} Thus, the
Hamiltonian of the total system can be written as
\begin{equation}\label{eq.10}
\begin{split}
H=&\Delta_s a_s^\dag a_s + G_s( a_s b_1^\dag + a_s^\dag b_1)
\\
& + (\Omega_{1} + G_1 {|\alpha_1(t)|}^2) {b_1}^\dag b_1
\\
& + \sum_{\omega}g({\omega})[b^\dag(\omega) b_1 \exp({-i\omega t})+\rm h.c.],
\end{split}
\end{equation}
where $a_s$ ($a_s^{\dag}$) represents the annihilation (creation)
operator of the optical mode in cavity $B$, and $\omega_s$ is the
corresponding inherent frequency. Here $\Delta_s=\omega_s-\omega_d$ is
the detuning frequency of cavity $B$, and $\omega_d$ is the
frequency of the external driving field. Also, $G_s$ denotes the
coupling strength between the optical mode and the mechanical
mode. To compensate the effect induced by the chaotic feedback on the
quantum memory system, we take the detuning frequency as $\Delta_s
= \Omega_{1} + G_s {|\alpha_1(t)|}^2$. In the rotating {reference} frame with
the unitary matrix
\begin{equation}\label{eq.11}
U=\exp\left[-{i}\int^{t}_{0}{(G_1 {|\alpha_1(\tau)|}^2+\Omega_1)(b_1^\dag b_1 + a_1^\dag a_1)}d\tau\right],
\end{equation}
the effective system Hamiltonian can be represented by
\begin{equation}\label{eq.10}
{H}_{\rm eff} = G_s( a_s^{^\dag} b_1  + a_s b_1^{^\dag}) + \sum_{\omega} {\tilde{g}({\omega})}[b^\dag(\omega) b_1 e^{-i(\Omega_{1}-\omega)t} + \rm h.c.],
\end{equation}
where $\tilde{g}({\omega})=\sqrt{M}g({\omega})$ and $M$ is the
decoupling factor. After introducing the adiabatic approximation
to eliminate the cavity mode shown in {Ref.~\cite{QM2}}, we use
$\tilde{b}_1$ to denote the annihilation operator of the
mechanical mode and the quantum Langevin equation of the
optomechanical system can be simplified as
\begin{equation}\label{QSDE_M}
\frac{d\tilde{b}_1}{dt}=-\frac{\nu + \Gamma_{\!1}}{2}\tilde{b}_1 - \sqrt{\nu}\,{a_{d}}-\sqrt{\Gamma_{\!1}}\, b_{\rm in}(t),
\end{equation}
${a}_{d}$ denotes the optical field fed into cavity $B$. Let
${a}_{d}=\alpha_d+\tilde{a}_d$, where $\alpha_d$ and $\tilde{a}_d$
denote the classical part and the quantum fluctuation of the
optical mode. The fluctuation terms $\tilde{a}_d$ and $b_{\rm in}$
satisfy the relations: $\langle{\tilde{a}_d(t)
\tilde{a}_d^{\dag}(t^{'})}\rangle=\delta(t-t^{'})$,
$\langle{{b}_{\rm in}(t) {b}_{\rm
in}^{\dag}(t^{'})}\rangle=(n+1)\delta(t-t^{'})$, where
$n\left(\Omega_1\right)\approx k_B T/\hbar\Omega_1$ is the mean
thermal excitation phonon number. The parameter $\nu$ in
Eq.~(\ref{QSDE_M}) can be calculated by $\nu={(G_s
|\alpha_d|)^2}/{\gamma_s}$, where $G_s$ is the coupling strength
between the mechanical mode and the optical mode, and $\gamma_s$
is the damping rate of the optical mode~\cite{QM1}.

We now assume that the system is initially in a Gaussian state. We
use the fidelity $F_{\infty}$ between the initial state and the steady state of
the mechanical mode to characterize the efficiency of noise
decoupling, which can be calculated by~\cite{QM1}
\begin{equation}\label{eq.10}
\begin{split}
  F_{\infty}=&\langle{\Psi_{0}}|{\rho_{\infty}}|{\Psi_{0}}\rangle
 \\
 =&\prod_{j=\pm
 s}\left[\exp({j})+\frac{\Gamma_{\!1}(2n+1-\exp({j}))}{2(\nu+\Gamma_{\!1})}\right]^{-\frac{1}{2}}.
\end{split}
\end{equation}
Here $s$ is the squeezing factor (see Appendix C). The
steady-state fidelity $F_{\infty}$ mainly depends on four factors:
the mean thermal excitation phonon number $n$, the coupling
strength $\nu$, the squeezing factor $s$, and the mechanical
damping rate $\Gamma_{\!1}$.
We can see that the fidelity $F_{\infty}$ can be increased by
decreasing the mechanical damping rate $\Gamma_{\!1}$, and, as shown
in Sec.~\Rmnum{2}, $\Gamma_{\!1}$ can be {reduced} by introducing {a}
chaotic feedback loop. In fact, after introducing the chaotic
feedback control, the effective damping rate of the mechanical
mode {is given by}
\begin{equation}\label{eq.11}
\Gamma_1^{\prime}=M\Gamma_{\!1}.
\end{equation}
Thus the modified fidelity $F_{\infty}^{\prime}$ can be written as
\begin{equation}\label{eq.10}
  F_{\infty}^{\prime} =\prod_{j=\pm s}\left[\exp({j})+\frac{\Gamma_1^{\prime}(2n+1-\exp({j}))}{2(\nu+\Gamma_1^{\prime})}\right]^{-\frac{1}{2}}.
\end{equation}
When the controller in the feedback loop enters the chaotic
regime, we have $\Gamma_1^{\prime}\approx0$, and thus
$F_{\infty}^{\prime}\approx1$, which means almost perfect quantum
state transfer.

Then, we numerically calculate the steady-state fidelity
$F_{\infty}$ between the input state and the steady state of the
mechanical resonator. Two different Gaussian input states are
considered: coherent states and squeezed states.

\subsection{\label{sec:level3}Coherent input state}
In this subsection, we consider the quantum mechanical memory
system with a coherent input state. For a coherent input
state, the squeezing factor $s=0$. Thus, in this case, the fidelity
{$F_{\infty}^c$} can be simplified as
\begin{equation}\label{F_C}
  F_{\infty}^{c}=\left[1+\frac{\Gamma_{\!1} n}{\nu+\Gamma_{\!1}}\right]^{-1}.
\end{equation}

\begin{center}
\begin{figure}[ht]
\centering
\includegraphics[width=3.5in]{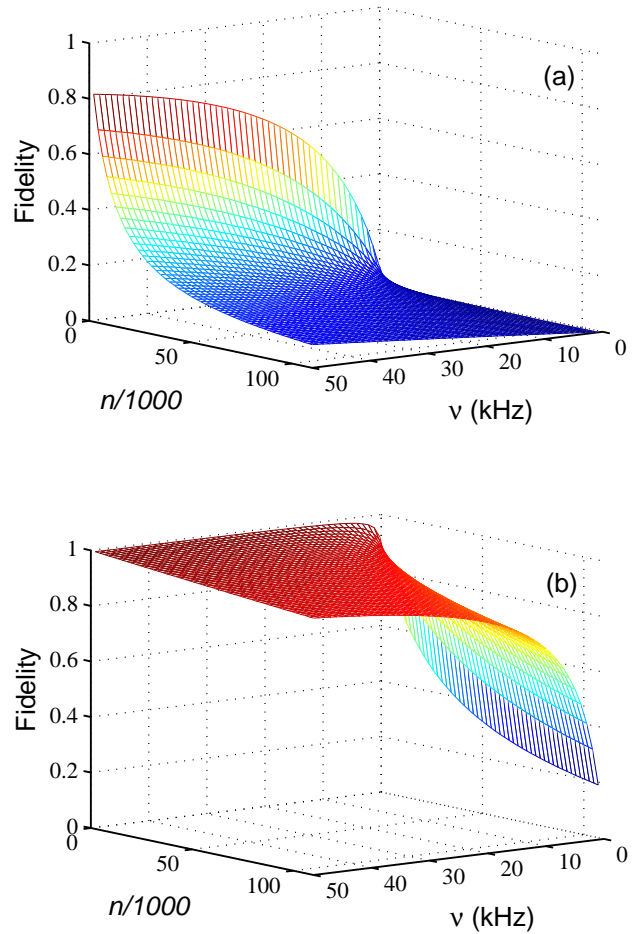}
\caption{(Color online) The fidelity (a) {before} the noise
decoupling and (b) {after} the noise decoupling.
{Here $n$ is the mean thermal excitation phonon
number which follows the Boltzmann distribution, and $\nu$ is a
parameter related to the optomechanical coupling strength.} {The
parameters are}: $\Omega_1/2\pi=1$ MHz,
$\Gamma_{\!1}/2\pi=5$ Hz for (a), and $\Gamma_1^{\prime}/2\pi=0.037$ Hz for
(b).} \label{coherent_state}
\end{figure}
\end{center}

By comparing the fidelity between the {input state}
and the steady state of the mechanical mode (under the
noise-decoupling control [see Fig.~\ref{coherent_state}(a)] and
without the noise-decoupling control [see
Fig.~\ref{coherent_state}(b)]), we find remarkable improvement of
the efficiency of the quantum memory by introducing chaotic
control. From Fig.~\ref{coherent_state}(a) and
Fig.~\ref{coherent_state}(b), we can observe {that}
 {the decrease of} the mean thermal excitation phonon number $n$ or
 {the increase of} the parameter $\nu$ {would} lead
to the improvement of the fidelity of the quantum transfer.
{If we fix the parameter $\nu=50$ kHz, the fidelity
of the quantum transfer will fall to zero rapidly when
increasing the excitation phonon number $n$ without introducing
the noise-decoupling control [Fig.~\ref{coherent_state}(a)]. We
find that the fidelity of the quantum {memory} is
increased and approaches one even when the mean thermal
excitation phonon number $n$ exceeds $10^5$ after introducing
the noise-decoupling control.} This means that our noise-decoupling
method efficiently reduces the damping rate of the mechanical mode
$\Gamma_{\!1}$, and thus protects the coherent input state
{from decoherence}.
\subsection{\label{sec:level3}Squeezed input state}
\begin{center}
\begin{figure}[ht]
\centering
\includegraphics[width=3.5in]{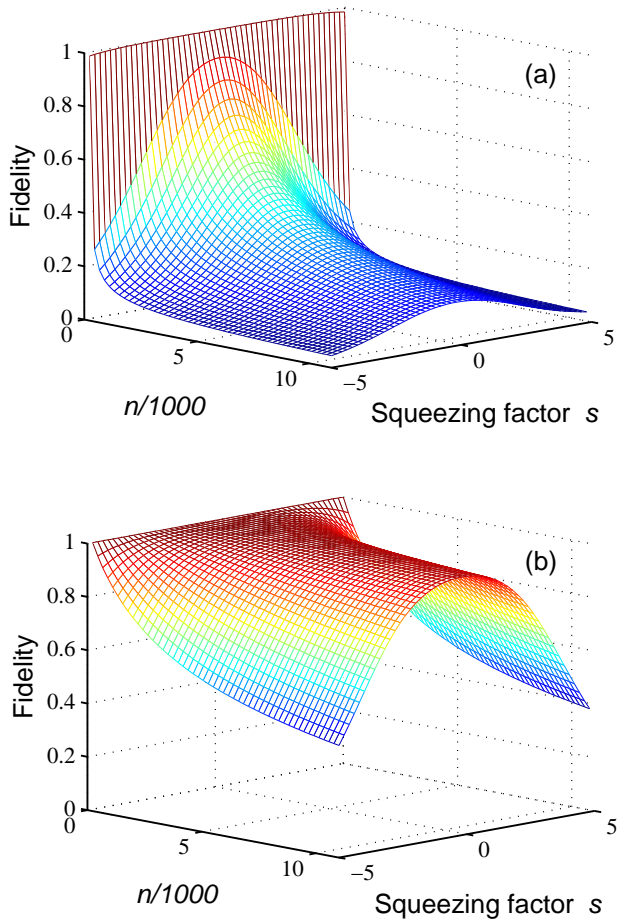}
\caption{(Color online) The fidelity (a) {before} noise decoupling
and (b) {after} noise decoupling. The parameters used are:
$\nu/2\pi=10~\rm kHz$,
$\Gamma_1/2\pi=5~\rm Hz$ for (a), and
$\Gamma_1^{\prime}/2\pi=0.037~\rm Hz$ for (b). The natural
frequency of the mechanical mode is assumed as
$\Omega_1/2\pi=1~\rm {MHz}$. } \label{squeez_state}
\end{figure}
\end{center}

Let us consider the case that the input state is a squeezed state
with squeezing factor $s\neq0$. By adjusting the squeezing factor
$s$ and the mean thermal excitation phonon number $n$, we
study the fidelity between the input squeezed state and the steady
state of the mechanical mode.

Compared to case without noise-decoupling control shown in
Fig.~\ref{squeez_state}(a), the fidelity under noise-decoupling
control is significantly improved [see Fig.~\ref{squeez_state}(b)] for different chosen system parameters. As shown in
Fig.~\ref{squeez_state}(a) and (b), the fidelity decreases when {increasing the squeezing factor $s$} and the mean
thermal excitation phonon number $n$. Here we vary the squeezing
factor $s$ from $-5$ to $5$, and it can be found that the curve of
fidelity is symmetrical about the plane $s=0$ in the
three-dimensional fidelity space. For each parameter $n$, the
fidelity is maximized when $s=0$, which corresponds to the case
that the input state is a coherent state. The quantum information
stored in the memory system is more likely to be damaged by the
heat bath {when increasing the degree of the} squeezing factor $s$. As shown
in Fig.~\ref{squeez_state}, the fidelity of quantum transfer
$F=0.16$ is very low when $n=10^5$ and $s=0$ without the
noise-decoupling control [see Fig.~\ref{squeez_state}(a)], while,
with the same condition, the fidelity is enhanced to be $F=0.96$
if we introduce the noise-decoupling control [see
Fig.~\ref{squeez_state}(b)]. When the squeezing factor $s$ is
increased to approach $5$, the fidelity decreases to zero rapidly
without the noise-decoupling control [see Fig.~\ref{squeez_state}(a)], while it will remain nonzero, i.e., $F=0.38$, when we
introduce the noise-decoupling control [see
Fig.~\ref{squeez_state}(b)].

\section{Conclusion}
To summarize, by introducing a chaotic feedback control
loop, we propose a strategy to decouple a nanomechanical
resonator in a quadratically-coupled optomechanical system from
the environmental noises. The main advantage of this method is to introduce a chaotic
controller to significantly broaden the spectrum of a mechanical
resonator and thus efficiently suppress the environmental noise.
As an application, we study this proposed the noise-decoupled nanomechanical
resonator under chaotic coherent feedback control as a quantum
memory to store the information transferred from external optical
signals. Two different input states, i.e., {coherent and
squeezed states}, are studied to show the efficiency of the
quantum memory. {The numerical results show that the fidelity of the quantum memory has been greatly improved after introducing our noise-decoupling strategy.} We believe that this nonlinear coherent feedback
strategy will have various applications, such as nonlinear
modulation of photon transport and high-sensitivity quantum
measurements, which will be considered in future work.

\begin{acknowledgments}
N.Y. would like to thank Dr. Yong-Chun Liu for his constructive suggestions, and the useful discussions with Hao-Kun Li and Dr. Sahin Kaya Ozdemir are also acknowledged. J.Z. and R.B.W. are supported by the National Natural Science Foundation of China (NSFC)
under Grants No. 61174084, No. 61134008, and No. 60904034. Y.X.L.
is supported by the National Natural Science Foundation of China
under Grants No. 10975080 and No. 61025022. Y.X.L. and J.Z. are
supported by the National Basic Research Program of China (973
Program) under Grant No. 2014CB921401, the Tsinghua University
Initiative Scientific Research Program, and the Tsinghua National
Laboratory for Information Science and Technology (TNList)
Cross-discipline Foundation. J.Z. is also partially supported by
Open Project of State Key Laboratory of Robotics. C.W.L. is
supported by the NSFC
under Grants No. 61174068. F.N. is supported by the RIKEN iTHES
Project, MURI Center for Dynamic Magneto-Optics, and
Grant-in-Aid for Scientific Research (S).

\end{acknowledgments}

\appendix
\section{\label{sec:level1}Theory of Markovian coherent feedback network\protect\\ }
To study the multi-channel quantum input-output
network, we now introduce the $SLH$ {method} presented in Ref.~\cite{SLH}. In the $SLH$ language, an open quantum system
can be fully characterized by $G=(S,L,H)$, where $S$ denotes a
$n\times n$ unitary scattering matrix, which satisfies
$SS^{\dag}=S^{\dag}S=I$, $L$ represents the dissipation operator
which is determined by the dissipation channels induced by the
input fields, and $H$ is the free Hamiltonian of the system. Within the framework of $G=(S,L,H)$, the quantum Langevin equation of an arbitrary system operator $X$ is given by

\begin{equation}\label{eq.5}
\begin{split}
\dot{X} =& -i[X, H_{\rm sys}]+ \{L^\dag [X, L] + [L^\dag, X]L\}/2
\\
& + \{b_{\rm in}[L^\dag, X] + [X,L]b_{\rm in}^\dag\}.
\end{split}
\end{equation}

\begin{figure}[ht]
\includegraphics[width=\textwidth,bb=0 0 900 100]{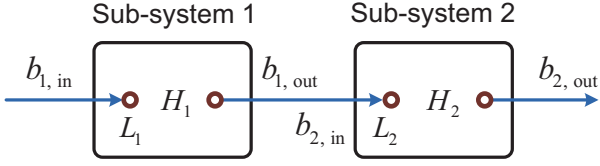}
\caption{ (Color online) Schematic diagram of the series product
of two cascaded-connected components.} \label{2casecade}
\end{figure}

The $SLH$ {method} provides a convenient way to study the all-optical
quantum coherent forward and feedback networks~\cite{SLH}. For
example, we show in Fig.~\ref{2casecade} two quantum components:
$G_{1}=(S_{1},L_{1},H_{1})$ and $G_{2}=(S_{2},L_{2},H_{2})$. The
series product of these two components can be parameterized by

\begin{eqnarray}
G_{2}\triangleright G_{1}&=&[S_{2}S_{1},L_{2}+S_{2}L_{1}, \nonumber\\
& &
H_{1}+H_{2}+\frac{1}{2i}(L_{2}^{\dag}
S_{2}L_{1}-L_{1}^{\dag}S_{2}^{\dag}L_{2})]. ~\label{ecuacion}
\end{eqnarray}

\begin{figure}[ht]
\includegraphics[width=\textwidth,bb=-10 -10 800 250]{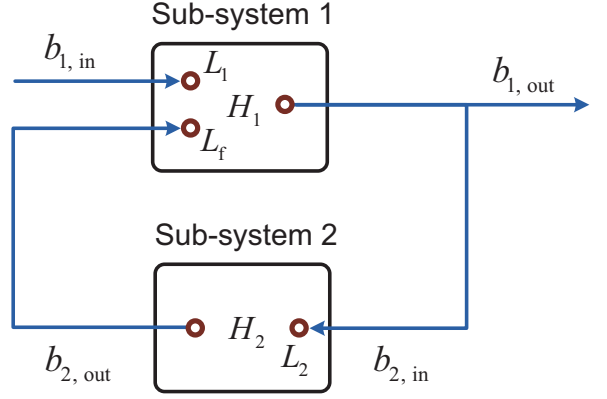}
\caption{ (Color online) Schematic diagram of a coherent feedback loop.}
\label{feedback_loop}
\end{figure}

A typical coherent feedback control system is shown in
Fig.~\ref{feedback_loop}, which is composed of the controlled
system, i.e., system $1$, and the controller, i.e., system
$2$. This coherent feedback control system can be seen as a
series product of three components: $G_{1}=(S_{1},L_{1},H_{1})$,
$G_{2}=(S_{2},L_{2},H_{2})$, and $G_{f}=(S_{f},L_{f},H_{1})$.
Thus, the corresponding $SLH$ parameters of this feedback control
system can be represented by
\begin{eqnarray}\label{G_feedback}
G_{f}\triangleright G_{2}\triangleright G_{1} &=& (S,L,H_{\rm sys}),
\end{eqnarray}
where
\begin{subequations}\label{G_feedback}
\begin{equation}
S=S_fS_2S_1,\ \ \ \ L=S_2S_1L_{1}+S_1L_{2}+L_{f},
\end{equation}
\begin{equation}
H_{\rm sys}=H_{1}+H_{2}+H_{\rm int},
\end{equation}
\end{subequations}
and the interaction Hamiltonian induced by the coherent feedback
loop is given by
\begin{equation}\label{H_int_S}
\begin{split}
H_{\rm{int}}=&\frac{1}{2i}(L_2^\dag S_2 L_1-L_1^\dag S_2^{\dag} L_2
\\&+L_{f}^\dag S_f L_2-L_2^\dag S_f^{\dag} L_{f}+L_{f}^\dag S_f S_2 L_1-L_1^\dag S_2^{\dag} S_f^{\dag} L_{f}).
\end{split}
\end{equation}
As an example, let us consider our feedback-induced
noise-decoupling system. As introduced in section $3$, a quadratically-coupled optomechanical device (system $1$)
and a chaotic controller (system $2$) are connected by optical
fields to construct a coherent feedback loop, which is similar to
that given by Eq.~(\ref{G_feedback}). Let $a_1$ ($a_2$) be the
annihilation operator of the cavity mode in quantum system $1$
($2$) with corresponding damping rate $\gamma_1$ ($\gamma_2$), and
$\gamma_f$ is the damping rate of the controlled cavity induced by
the feedback field. In this case, we have
$L_1=\sqrt{\gamma_1}\,a_1$, $L_2=\sqrt{\gamma_2}\,a_2$, and
$L_f=\sqrt{\gamma_f}\,a_1$, and $S_{1}=S_{2}=S_{f}=I$. From
Eq.~(\ref{H_int_S}), the dissipation operator of the total
feedback loop can be written as
\begin{equation}\label{L}
L=(\sqrt{\gamma_1}+\sqrt{\gamma_f})a_1+\sqrt{\gamma_2}\,a_2,
\end{equation}
and the total Hamiltonian of the quantum feedback loop can be
obtained from Eq.~(\ref{G_feedback}) and Eq.~(\ref{H_int_S}) as
\begin{equation}
\begin{split}
H_{\rm sys}&=H_1 + H_2 + H_{\rm int}
\\&=H_1 + H_2 + \frac{1}{2i}(\sqrt{\gamma_1 \gamma_2}-\sqrt{\gamma_2 \gamma_f})(a_2^\dag a_1-a_1^{\dag} a_2).
\end{split}
\end{equation}

Accordingly, the quantum Langevin equations of the two cavity
modes $a_1$ and $a_2$ can be represented by
\begin{subequations}
\label{Langevin3}
\begin{equation}
\begin{split}
\dot{a}_1=&-i[a_1, H_{1}+H_2] - \frac{1}{2}(\sqrt{\gamma_1}+\sqrt{\gamma_f})^{2} a_1\\
& -\sqrt{\gamma_2\gamma_f}\,a_2-(\sqrt{\gamma_1}+\sqrt{\gamma_f})\,a_{1,\rm in},\label{subeq:1}
\end{split}
\end{equation}
\begin{equation}
\begin{split}
\dot{a}_2=-i[a_2, H_{1}+H_2] - \frac{\gamma_2}{2}
a_2-\sqrt{\gamma_1 \gamma_2}\,a_1-\sqrt{\gamma_2}\,a_{2,\rm
in},\label{subeq:2}
\end{split}
\end{equation}
\end{subequations}
where $a_{1,\rm in}$ ($a_{2,\rm in}$) is the input field fed into
the system $1$ ($2$). By substituting Eqs.~(\ref{H_1}) and
(\ref{eq.2}) into Eq.~(\ref{Langevin3}), we can obtain the quantum
Langevin equation given by Eq.~(\ref{Langevin1}).

\section{Derivation of the decoupling coefficient $M$}
The decoupling coefficient $M$ is determined by the classical cavity field $f(t)$, which can be decomposed into a series of frequency components by the Fourier transform~\cite{DC1,APS1}.
\begin{equation}
f(t)=\sum^{\infty}_{n=0}A_{n} \cos(\omega_{n}t + \varphi _{n}),
\end{equation}
where $\omega_{n}$, $A_{n}$ and $\varphi_{n}$ denote the
frequency, the amplitude and the initial phase of the $n$-th
frequency components. Integrating $f(t)$ gives the
control-induced phase shift
\begin{equation}\label{phase}
\theta (t)=\int_{0}^{t}\!\!f(\tau)\;d\tau =
\sum_{n=0}^{\infty}\frac{A_{n}}{\omega_{n}} \sin (\omega_{n}t +
\varphi _{n}).
\end{equation}
By introducing the Bessel-series expansion, we have
\begin{equation}\label{phase2}
\begin{split}
\exp\left[-i\theta (t)\right]&=\exp
\left[-i\sum_{n=0}^{\infty}\frac{A_{n}}{\omega_{n}} \sin
(\omega_{n}t + \varphi _{n})\right]
\\
&=\prod_\alpha \sum_{n} J_{n \alpha}
\left(\frac{A_{\alpha}}{\omega
_{\alpha}}\right)\exp[-in_{\alpha}\omega_\alpha t
-in_{\alpha}\varphi_{\alpha}],
\end{split}
\end{equation}
where $J_{n \alpha}$ is the $n$-th Bessel function of the first
kind. We then neglect the high-order terms in Bessel series, which
can be considered as the fast variables in the system, and only
keeps the zero-order terms in Eq.~(\ref{phase2}), by which we have
\begin{equation}\label{e}
\overline{\exp(-i\theta (t))}=\prod_\alpha J_{0} \left(\frac{A_{\alpha}}{\omega _{\alpha}}\right)=\exp \left[\sum_{\alpha}\ln{J_0}\left(\frac{A_{\alpha}}{\omega_{\alpha}}\right) \right].
\end{equation}
Under the condition that $A_{\alpha}\ll\omega_{\alpha}$, the
zero-order Bessel term can be approximately expressed as
$J_{0}({A_{\alpha}}/{\omega _{\alpha}}) \approx
1-{({A_{\alpha}}/2{\omega _{\alpha}})^2}$. Furthermore, from
$A_{\alpha}\ll\omega_{\alpha}$, we have
$\ln(1-({A_{\alpha}}/2{\omega _{\alpha}})^2)\approx
-({A_{\alpha}}/2{\omega _{\alpha}})^2$. Thus Eq.~(\ref{e}) can be
simplified as
\begin{eqnarray}\label{J}
\prod_\alpha J_{0}
\left(\frac{A_{\alpha}}{\omega_{\alpha}}\right)&=&\exp{\left[-\frac{1}{4}\sum_{\alpha}{\frac{A_{\alpha}^2}{\omega_{\alpha}^2}}\right]} \nonumber
\\
&=&\exp{\left[-\frac{\pi}{2}\int_{\omega_l}^{\omega_u}{\frac{S_{f}(\omega)}{\omega_{\alpha}^2}d\omega}\right]}.
\end{eqnarray}
Let $\sqrt{M}=\overline{\exp(-i\theta (t))}$, and $M$ is defined as
the decoupling factor, then from Eq.~(\ref{J}) we have
\begin{eqnarray}
M=\exp{\left[-{\pi}\int_{\omega_l}^{\omega_u}{\frac{S_{f}(\omega)}{\omega_{\alpha}^2}d\omega}\right]}.
\end{eqnarray}

\section{\label{sec:level1}fidelity of the quantum memory\protect\\ }
The Langevin equation of the mechanical operator $\tilde{b}_1$ is
shown in Eq.~(\ref{QSDE_M}). The steady value of the mechanical
mode can be obtained by setting ${d\tilde{b}_1}/{dt}=0$ as
\begin{equation}\label{eq.10}
\langle{{\tilde{b}_1}(\infty)}\rangle=\frac{-2\sqrt{\nu}}{\nu+\Gamma_{\!1}}\alpha_d,
\end{equation}
where $\langle\cdot\rangle$ is the average over the input vacuum
fluctuation. We then define the quantum Wiener processes
$A(t)=\int_0^{t}{\tilde{a}_d (t^{'})}d{t^{'}}$,
$B(t)=\int_0^{t}{{b_{\rm in}} (t^{'})}d{t^{'}}$, by which we can
obtain the quantum stochastic differential equation from
Eq.~(\ref{QSDE_M}) as
\begin{equation}\label{eq.10}
d{\tilde{b}}_1=-\frac{\nu + \Gamma_1}{2}\,\tilde{b}_1{dt} -
\sqrt{\nu}\,\alpha_d{dt}-\sqrt{\nu}\,d\!A-\sqrt{\Gamma_1} d\!B.
\end{equation}
The quantum fluctuation terms $d\!A$ and $d\!B$ satisfy that
\begin{eqnarray}\label{eq.5}
\langle d\!A\rangle=\langle d\!B \rangle=0,
\end{eqnarray}
and obey the quantum Ito rules
\begin{eqnarray}\label{Ito}
&&\!\!\!\!\!\!\!\!\!\!\!\!\!\!\!\! d\!A\;d\!A^{\dag}=(N+1)dt,\ \ \ d\!A^{\dag}d\!A=N\,dt,\nonumber
\\&&\!\!\!\!\!\!\!\!\!\!\!\!\!\!\!\! (d\!A)^2=M\,dt,\ \ \ (d\!A^{\dag})^2=M^{\dag}dt,
\\&&\!\!\!\!\!\!\!\!\!\!\!\!\!\!\!\! d\!B\;d\!B^{\dag}=(n+1)dt,\ \ \ d\!B^{\dag}d\!B=n\,dt, \nonumber
\end{eqnarray}
{where $n$ represents the thermal excition  number, $N$ is the effective photon number, and $M$ denotes the squeezing parameter. Here $M$ and $N$ satisfy the inequality $M^2\geq N(N+1)$.}
{Then we introduce the squeezing factor $s$~\cite{QO} of the input quantum state, which is given by}
\begin{eqnarray}\label{squeezing_factor}
s=\ln{\left[M+M^{*}+2N+1\right]}.
\end{eqnarray}

To calculate the fidelity of the quantum memory, let us define the
normalized position $x=(\tilde{b}_1+\tilde{b}^{\dag}_1)/\sqrt{2}$,
momentum $p=(\tilde{b}_1-\tilde{b}^{\dag}_1)/\sqrt{2}i$, and the
conjugate vector $\textbf{z}=(x,p)$ of the mechanical mode. We
also introduce the symmetrized covariance matrix $V$, which is
given by
\begin{equation}\label{eq.10}
 V=\frac{1}{2}[\Delta \textbf{z}\ \Delta \textbf{z}^{\textrm{T}}+(\Delta \textbf{z}\ \Delta \textbf{z}^{\textrm{T}})^{\textrm{T}}],
\end{equation}
where $\Delta \textbf{z} =\textbf{z}-\langle \textbf{z}\rangle$.
With Ito's rule $d(ab)=(da)b+a(db)+da\,db$, the time evolution
of the covariance matrix $V$ is described by the Lyapunov
differential equation
\begin{equation}\label{V_stable}
 \dot{V}=AV+VA^{T}+\Gamma_1(n+1/2)I_{2}+\nu\,\Lambda,
\end{equation}
where $A=-\left[(\nu+\Gamma_1)/2\right]I_2$, and $I_2$ is the
two-dimensional identity matrix. Here, $\Lambda$ is a matrix related to
the degree of squeezing, which can be calculated by
\begin{equation}
 \Lambda= \frac{1}{2}
\left(
\begin{array}{cc}
{2N+1+M+M^{*}} & {M-M^{*}}\\
{M-M^{*}} &  {2N+1-(M+M^{*})}
\end{array}
\right).
\end{equation}
For a squeezed input state, the fidelity between the initial state
and the steady state of the mechanical mode is given by
\begin{equation}\label{C9}
\begin{split}
  F_{\infty}=&\langle{\Psi_{0}}|{\rho_{\infty}}|{\Psi_{0}}\rangle=\frac{1}{\sqrt{\det(V_{\infty}+V_{0})}}
 \\
 =&\prod_{j=\pm s}\left[\exp{(j)}+\frac{\Gamma_1(2n+1-\exp{(j)})}{2(\nu+\Gamma_{\!1})}\right]^{-\frac{1}{2}},
\end{split}
\end{equation}
where $V_{\infty}$ denotes the stationary solution of the Lyapunov
differential equation [Eq.~(\ref{V_stable})] and $V_0$ is the
covariance matrix of the input state, which can be calculated by
\begin{equation}
V_0=\frac{1}{2}\left(
\begin{array}{cc}
 \exp{(s)} & 0\\
0 & \exp{(-s)}
\end{array}
\right).
\end{equation}
When the input state is a coherent state, such that $M\,=\,N\,=\,0$
and thus $s\,=\,0$, the fidelity in this case can be simplified as
\begin{equation}\label{C11}
\begin{split}
  F_{\infty}^{c}=\frac{1}{\sqrt{\det(V_{\infty}+V_{0})}}
 \\
 =\left[1+\frac{\Gamma_{\!1} n}{\nu+\Gamma_{\!1}}\right]^{-1}.
\end{split}
\end{equation}
It can be found from Eqs. (\ref{C9}) and (\ref{C11}) that the fidelity increases when increasing the mechanical damping rate $\Gamma_1$ for both squeezed states and coherent states.

\end{document}